\documentclass[aip,jcp,reprint]{revtex4-1}

\usepackage{amsmath}
\usepackage{hyperref}
\usepackage{cleveref}

\IfFileExists{.git/HEAD}{
    \usepackage{tikz}
    \usepackage{pgfplots}
    \pgfplotsset{compat=1.5.1}
    \usetikzlibrary{external}
    \usepgfplotslibrary{external}
    \tikzexternalize[prefix=tikz/]

    \usepackage{todonotes}

}{
    \usepackage{tikzexternal}
    \tikzexternalize
\IfFileExists{tikz/\jobname-figure0.pdf}{
    \tikzsetexternalprefix{tikz/}
}{
    \relax
}
    \let\tikzexternaldisable\relax
    \let\tikzexternalenable\relax
    \newcommand{\todo}[2][]{}
}

\makeatletter
\def\input@path{{./}{pictures/}}
\makeatother
\graphicspath{{./}{pictures/}}

\usepackage{jabbrv}
\DefineSpuriousJournalWord{The}
\DefineSpuriousJournalWord{on}
\DefineJournalException{Chemistry: A European Journal}{Chem. Eur. J.}
\DefineJournalException{Journal of Physics: Condensed Matter}{J. Phys. Condens. Matter}
\DefineJournalException{Angewandte Chemie}{Angew. Chem.}
\DefineJournalException{Computers \& Fluids}{Comput. Fluids}
\DefineJournalAbbreviation{Surfaces}{Surf}
\DefineJournalAbbreviation{Reports}{Rep}
\DefineJournalAbbreviation{Physique}{Phys}

\usepackage[exponent-product=\cdot]{siunitx}

\makeatletter
\@ifpackageloaded{tikz}{
    \renewcommand{\todo}[2][]{\tikzexternaldisable\@todo[#1]{#2}\tikzexternalenable}
}{
    \relax
}
\makeatother

\usepackage{bm}
\renewcommand{\vec}[1]{\bm{#1}}
\newcommand{\unitvec}[1]{\hat{\bm{#1}}}
\newcommand{\agrid}{a_\text{grid}}
\newcommand{\tgrid}{\tau}

\begin{document}


\title{A Lattice Boltzmann Model for Squirmers}

\author{Michael Kuron}
\email{mkuron@icp.uni-stuttgart.de}
\affiliation{Institute for Computational Physics, University of Stuttgart, Allmandring 3, 70569 Stuttgart, Germany}

\author{Philipp St\"ark}
\affiliation{Institute for Computational Physics, University of Stuttgart, Allmandring 3, 70569 Stuttgart, Germany}

\author{Christian Burkard}
\affiliation{Institute for Computational Physics, University of Stuttgart, Allmandring 3, 70569 Stuttgart, Germany}

\author{Joost de Graaf}
\affiliation{Institute for Theoretical Physics, Center for Extreme Matter and Emergent Phenomena, Utrecht University, Princetonplein 5, 3584 CC Utrecht, The Netherlands}

\author{Christian Holm}
\affiliation{Institute for Computational Physics, University of Stuttgart, Allmandring 3, 70569 Stuttgart, Germany}

\date{\today}


\begin{abstract}
The squirmer is a simple yet instructive model for microswimmers,
which employs an effective slip velocity on the surface of a spherical swimmer to describe its self-propulsion.
We solve the hydrodynamic flow problem with the lattice Boltzmann (LB) method,
which is well-suited for time-dependent problems involving complex boundary conditions.
Incorporating the squirmer into LB is relatively straight-forward,
but requires an unexpectedly fine grid resolution to capture the physical flow fields and behaviors accurately.
We demonstrate this using four basic hydrodynamic tests:
Two for the far-field flow---accuracy of the hydrodynamic moments and squirmer-squirmer interactions---and
two that require the near field to be accurately resolved---a squirmer confined to a tube and one scattering off a spherical obstacle---which
LB is capable of doing down to the grid resolution.
We find good agreement with (numerical) results obtained using other hydrodynamic solvers
in the same geometries
and identify a minimum required resolution to achieve this reproduction.
We discuss our algorithm in the context of other hydrodynamic solvers
and present an outlook on its application to multi-squirmer problems.
\end{abstract}

\maketitle


\section{Introduction}
\label{sec:intro}

Directed motion, or motility, is of paramount importance to biology\cite{cates12a}.
For example, it allows bacteria to move toward a food source\cite{hoell11a} and
fish to swim in formations that protect them from predators\cite{marchetti13a}.
In the presence of other bacteria,
motility can lead to collective effects\cite{kearns10a,wensink12a,gachelin14a} bearing resemblance of the schooling of fish or the swarming of birds,
suggesting that the specifics of the propulsion, or even the length scale on which it occurs,
have little effect on the overall behavior.
Yet we know that these two situations are drastically different
from the perspective of interactions via the medium,
which are tied to the way the organisms achieve propulsion\cite{marchetti13a}.

To experimentally better understand how the propulsion method affects the motion of an individual biological swimmer
and how the motion affects the collective behavior of many biological swimmers together,
artificial analogs have been developed.
Realizations include catalytic\cite{howse07a,erbe08a} and self-thermophoretic\cite{jiang10c,buttinoni12a} propulsion methods.
These models have a well-defined geometrical shape and characterizable chemical properties,
thus eliminating biological complications like shape changes or the beating of cilia.
Yet, despite their simplicity,
they show the same kinds of collective effects as their biological counterparts\cite{ibele09a,palacci13a}.

Theoretical description of motility and the associated out-of-equilibrium phenomena is possible
using models such as the one by Vicsek\cite{vicsek95a} or the active Brownian particle (ABP) model\cite{romanczuk12a,ebeling99a}.
These approaches have been quite successful in qualitatively capturing the behaviors observed in nature.
However both neglect the hydrodynamic interactions mediated by the surrounding fluid,
which can be important for microorganisms and their artificial counterparts\cite{marchetti13a}.
One way to overcome this limitation is the squirmer model\cite{lighthill52a,blake71b}:
here, the microswimmer is described as a spherical object with a simple inhomogeneous surface slip velocity, typically suspended in a Newtonian fluid.
The squirmer model's long-ranged hydrodynamic interactions
lead to reorientation like in the Vicsek model,
and, when complemented with a near-field repulsion,
it accounts for the collisions that are captured by the ABP model.
The squirmer model has proven to be an effective tool to model the effect of hydrodynamics in suspensions of both bacteria and man-made swimmers\cite{alarcon13a,gonnella15a,zoettl14a}.


In this paper, we implement the squirmer model numerically
using the lattice Boltzmann (LB) method\cite{mcnamara88a},
making use of the \citeauthor{ladd94a} moving boundary conditions\cite{ladd94a}.
LB is in general a Navier-Stokes solver,
but it can serve as a Stokes solver at the Reynolds numbers relevant to the systems considered in the present paper.
The main advantages of LB over competing methods are momentum and mass conservation to machine precision,
very low compressibility and good obedience of the Stokes regime,
as well as facile coupling to suspended particles.
Furthermore, the algorithm scales to parallelize across large supercomputers\cite{godenschwager13a} and is fully deterministic.
Previous simulational studies of squirmers have used methods such as
multi-particle collision dynamics\cite{downton09a,theers16a,theers18a} (MPCD),
finite element method\cite{zhu12b,aguillon12a,decorato17a} (FEM),
boundary element method\cite{ishikawa06a,ishimoto13a,zhu13a,uspal15b} (BEM),
and Stokesian dynamics\cite{ishikawa08a,ishikawa08b,chamolly17a} (SD),
but also LB\cite{lintuvuori16a,llopis10a,alarcon13a,alarcon17a}.
We verify our implementation against well-known results from the literature\cite{ishikawa06a,zhu13a,spagnolie15a}
and show that even for these basic cases several new things may be learned.
Specifically, we study squirmers in bulk both individually and scattering off each other,
as well as a squirmer oscillating between the two walls of a channel.
Finally, we consider the interaction between a squirmer and an immobile spherical obstacle.
We find that LB is well-capable of reproducing results obtained from other methods\cite{ishikawa06a,zhu13a},
but requires a higher resolution to reproduce accurate results
than is known from experience with passive particles in LB.
This insight will enable us to accurately simulate dense suspensions of squirmers in the future
and to study phenomena such as motility-induced phase separation\cite{cates15a,gonnella15a,zoettl14a,theers18a}.

The remainder of this paper is laid out as follows:
In \cref{sec:model}, we discuss the squirmer model.
In \cref{sec:lb}, we introduce the relevant aspects of the LB method.
In \cref{sec:farfield,sec:nearfield}, we apply this numerical method to squirmers in bulk and interacting with boundaries, respectively.
There, we also discuss implementation problems that arise
and how our simulations compare to previous implementations,
before we conclude in \cref{sec:conclusion}.

\section{The Squirmer Model}
\label{sec:model}

The near-field flow of a microswimmer is highly dependent on the specifics of its propulsion\cite{drescher10a},
so theoretical descriptions tend to resort to simple models capable only of producing the correct far-field behavior.
One such model is the squirmer, initially introduced by \citet{lighthill52a} to explain swimming by oscillatory shape change.
\citet{blake71b} later adapted it to describe the envelope of the ciliar motion of \emph{Paramecium}, a microorganism.
Both authors expand the flow around a spherical swimmer into spherical harmonics
and discover that only the first two modes are needed to accurately capture the far-field flow.
Most current squirmer applications further do away with any flow radially through the surface\cite{ishikawa06a},
such that the ciliar motion on the surface of a sphere with radius $R$ corresponds to a flow velocity boundary condition\cite{blake71b}
\begin{equation}
\left.\vec{u}(\vec{r})\right|_{r=R}
=\left(B_1+B_2\frac{\unitvec{e}\cdot{\vec{r}}}{r}\right)
\left(\frac{\unitvec{e}\cdot\vec{r}}{r}\frac{\vec{r}}{r}-\unitvec{e}\right)
\label{eq:squirmer}
\end{equation}
in its rest frame,
where $\vec{r}$ is the position vector relative to the sphere's center, $B_n$ are constants, and $\unitvec{e}$ is the unit orientation vector of the sphere.
Due to the small size of a microswimmer,
a low Reynolds number approximation
\begin{equation}
\mathrm{Re}=\frac{2\rho v_0 R}{\eta} \ll 1,
\end{equation}
with the fluid density $\rho$ and dynamic viscosity $\eta$ and the characteristic velocity $v_0$
may be made to the Navier-Stokes equations that govern the fluid flow.
The resulting Stokes equations for hydrodynamics are given by
\begin{align}
\eta \nabla^2\vec{u}(\vec{r}) &= -\vec{\nabla}p(\vec{r}), \label{eq:stokes-p} \\
\vec{\nabla}\cdot\vec{u}(\vec{r}) &= 0,\label{eq:stokes-m}
\end{align}
where it is important to note that self-propulsion is force-free\cite{ishikawa09a,lauga09a}.
Here, $p$ refers to the pressure and
$\vec{\nabla}$, $\vec{\nabla}\cdot$ and $\nabla^2$ are the gradient, divergence and Laplace operators, respectively.
With the boundary condition of \cref{eq:squirmer}, one obtains the flow field\cite{blake71b,ishikawa06a}
\begin{align}
\vec{u}(\vec{r})=&\phantom{+}
B_1\frac{R^3}{r^3}\left(\frac{\unitvec{e}\cdot\vec{r}}{r}\frac{\vec{r}}{r}-\frac{1}{3}\unitvec{e}\right) \nonumber \\
&+ B_2\left(\frac{R^4}{r^4}-\frac{R^2}{r^2}\right)\left(\frac{3}{2}\left(\frac{\unitvec{e}\cdot\vec{r}}{r}\right)^2-\frac{1}{2}\right)\frac{\vec{r}}{r} \nonumber \\
&+ B_2\frac{R^4}{r^4}\frac{\unitvec{e}\cdot\vec{r}}{r}\left(\frac{\unitvec{e}\cdot\vec{r}}{r}\frac{\vec{r}}{r}-\unitvec{e}\right)
\label{eq:squirmerflow}
\end{align}
in the co-moving frame and sees that the squirmer moves with a velocity of
\begin{equation}
\vec{v}_0=\frac{2}{3}B_1\unitvec{e}
\label{eq:squirmer-speed}
\end{equation}
in the laboratory frame\cite{blake71b,ishikawa06a}.
The dipolarity
\begin{equation}
\beta=\frac{B_2}{B_1}
\end{equation}
is the ratio of the magnitudes of the second (force dipole, $r^{-2}$ decay) and first mode (source dipole, $r^{-3}$ decay).
Note that the former always brings along a source quadrupole term ($r^{-4}$ decay), which cannot be scaled independently.
$\beta$ describes the shape of the flow field and its sign distinguishes between different kinds of swimmers.
In the far field,
a pusher like the \emph{Escheria coli} bacterium\cite{drescher11a} ($\beta<0$) pushes away fluid at its front and back and pulls fluid in from its sides,
while pullers such as the \emph{Chlamydomonas reinhardtii} alga\cite{drescher10a} ($\beta>0$) pull fluid inward from the front and back and pushes it toward its sides.
\emph{Paramecium}\cite{ishikawa06b}, as a neutral swimmer with $\beta=0$, has a different far-field behavior and moves fluid from its front to its back.
The three types of squirmer are illustrated in \cref{fig:bulk}.


\section{Numerical Method}
\label{sec:lb}

To account for the above hydrodynamics numerically,
we employ the LB method\cite{mcnamara88a}.
Instead of solving the Stokes~\cref{eq:stokes-p,eq:stokes-m} directly,
this method solves the Boltzmann transport equation which obeys the same conservation laws
and describes the evolution of a function $f(\vec{r},\vec{v},t)$.
This is the system's single-particle phase space probability distribution,
giving the probability of finding a fluid molecule with velocity $\vec{v}$ at position $\vec{r}$ and time $t$.
The LB method linearizes the relaxation of $f$ to its Maxwellian equilibrium,
while discretizing space on a cubic lattice with lattice constant $\agrid$ and time in steps of $\tgrid$.
Only a finite set of velocities $\vec{c}_i$ is permitted,
specifically those that allow probability to flow between neighboring cells,
making the populations $f_i(\vec{r},t):=f(\vec{r},\vec{c}_i,t)$.
We choose the D3Q19 velocity set (3 dimensions and 18 face and edge neighbors).
Throughout this paper, we use the two relaxation time (TRT) collision operator\cite{ginzburg08a}
which relaxes symmetric and antisymmetric linear combinations of $f_i$ separately.
The symmetric relaxation corresponds to stress relaxation
with the relaxation time $\lambda_e$ determining the viscosity.
The antisymmetric relaxation gives a second relaxation time $\lambda_o$ as a free parameter
that can be used to improve the faithfulness of the boundary conditions\cite{ginzburg08b}.
The resulting LB equation is
\begin{align}
f_i(\vec{r}+\vec{c}_i\tgrid,t+\tgrid)
&=f_i(\vec{r},t) 
-\lambda_e(f_i^+-f_i^{\text{eq}+}) \nonumber \\
&\phantom{=f_i(\vec{r},t)}
-\lambda_o(f_i^--f_i^{\text{eq}-}) \label{eq:lb} \\
\intertext{with}
f_i^{\pm}(\vec{r},t) &= \frac{1}{2} \left( f_i(\vec{r},t) \pm f_{\bar{i}}(\vec{r},t) \right), \\
f_i^{\text{eq}\pm}(\vec{r},t) &= \frac{1}{2} \left( f_i^\text{eq}(\vec{r},t) \pm f_{\bar{i}}^\text{eq}(\vec{r},t) \right), \\
f_i^\text{eq}(\vec{r},t) &=  w_i\rho(\vec{r},t) \left( 1 + 3\vec{c}_i\cdot\vec{u}(\vec{r},t) \phantom{\frac{1}{6}} \right. \\
&\phantom{=} \left. + \frac{1}{6}\left(\vec{c}_i\cdot\vec{u}(\vec{r},t)\right)^2 -\frac{1}{6}u(\vec{r},t)^2 \right), \nonumber \\
\eta&= \rho\left(\frac{1}{3\lambda_e}-\frac{1}{6}\right), \\
\lambda_o\lambda_e&=\frac{3}{16},
\end{align}
where $f_i^{\pm}$ and $f_i^{\text{eq}\pm}$ are the symmetric ($+$) and antisymmetric ($-$) combinations of populations ($f_i$) and equilibrium populations ($f_i^\text{eq}$), respectively.
The index $\bar{i}$ is the one for which $-\vec{c}_i=\vec{c}_{\bar{i}}$.
From the populations, the macroscopic flow fields can be recovered:
\begin{align}
	\rho(\vec{r},t)&=\sum\limits_{i=1}^{19} f_i, \\
	\vec{u}(\vec{r},t)&=\sum\limits_{i=1}^{19} f_i\vec{c}_i. \label{eq:macroscopic-velocity}
\end{align}
Note that \cref{eq:lb,eq:macroscopic-velocity} do not need to be modified to account for a force\cite{guo02a,huang11b}
as there is no external (non-hydrodynamic) force applied to the fluid.
Fluid-particle interactions take place exclusively via boundary conditions.

Velocity boundary conditions such as the no-slip conditions we have on obstacles
are introduced by reflecting populations that stream into the boundary back into the fluid.
For non-zero velocity conditions, the reflected populations are shifted\cite{zou97a} to obtain
\begin{equation}
f_i(\vec{r}_\text{b}+\tgrid\vec{c}_i,t+\tgrid)
=f_{\bar{i}}(\vec{r}_\text{b},t)
+ \frac{6\rho w_i\tgrid^2}{\agrid^2} \vec{c}_i\cdot \vec{v}_\text{b}
\end{equation}
where $\vec{r}_b$ is a boundary node with velocity $\vec{v}_b$ and $\vec{r}_\text{b}+\tgrid\vec{c}_i$ is a fluid node.

If the boundary is allowed to move, the previous equation can be used as part of a particle coupling scheme introduced by \citet{ladd94a}.
A swimmer with its geometric center at $\vec{r}$ moving with velocity $\vec{v}$ and angular velocity $\vec{\omega}$ has a surface velocity of
\begin{equation}
\vec{v}_\text{b}(\vec{r}_\text{b},t)=\vec{v}(t) + \vec{\omega}(t) \times (\vec{r}_\text{b} - \vec{r}(t)) \label{eq:vb}
\end{equation}
in the lab frame.
To complete the particle coupling, one needs to account for the momentum transfer due to the reflection
by considering the force
\begin{align}
\vec{F}_\text{bb}(t)&=\agrid^3\sum\limits_{\vec{r}_\text{b}} \sum\limits_{i=1}^{19} \vec{c}_i \left(f_i(\vec{r}_\text{b},t)+f_{\bar{i}}(\vec{r}_\text{b}-\vec{c}_i\tgrid,t)\right) \label{eq:bbforce} \\
\intertext{and torque}
\vec{T}_\text{bb}(t)&=\agrid^3\sum\limits_{\vec{r}_\text{b}} \sum\limits_{i=1}^{19} \left(\vec{r}_\text{b}-\vec{r}\right) \times \vec{c}_i \left(f_i(\vec{r}_\text{b},t)\right. \nonumber \\
\phantom{\vec{T}_\text{bb}(t)}&\phantom{=\agrid^3\sum\limits_{\vec{r}_\text{b}} \sum\limits_{i=1}^{19} \left(\vec{r}_\text{b}-\vec{r}\right) \times \vec{c}_i(}
\left.+f_{\bar{i}}(\vec{r}_\text{b}-\vec{c}_i\tgrid,t)\right) \label{eq:bbtorque}
\end{align}
on the particle.
Unlike \citeauthor{ladd94a}'s original algorithm\cite{ladd94a},
we do not average $\vec{F}_\text{bb}(t)$ and $\vec{T}_\text{bb}(t)$ over two time steps.
This is generally only necessary if oscillations in these quantities are observed between consecutive time steps.
Note that the net force and torque of the system is still zero as required of a microswimmer\cite{purcell77a}:
the above force and torque just account for momentum transferred between fluid and particle.

One further detail of the moving boundary scheme is that cells transition between fluid and boundary over time.
\citet{aidun98a} suggested to set the populations of a cell to zero when it is converted to boundary
and to set the populations to their equilibrium value (based on the swimmer's velocity and the average density of surrounding fluid cells) when it is converted back to fluid.
This violates instantaneous mass conservation, but is unproblematic as average mass is conserved.
To conserve momentum, destruction and creation of populations at position $\vec{r}_\text{f}$
needs to be accounted for as a force
\begin{equation}
	\vec{F}_\text{c}(t)=\pm\frac{1}{\tgrid}\sum\limits_{i=1}^{19}f_i(\vec{r}_\text{f},t)\vec{c}_i \label{eq:consforce}
\end{equation}
on the swimmer.
Further enhancements of the moving-boundary method are reviewed in Ref.~\citenum{rettinger17a}.

In \cref{sec:orbiting},
we include a short-range repulsion between pairs of swimmers, as well as between swimmer-obstacle pairs,
in addition to the hydrodynamic forces of \cref{eq:bbforce,eq:consforce}.
This is a smooth approximation to a hard-core repulsion,
as introduced by \citet*{weeks71a}:
\begin{equation}
\vec{F}_\text{WCA}(r)=24\varepsilon\left(
\frac{2}{r}\left(\frac{\sigma}{r}\right)^{12}
-\frac{1}{r}\left(\frac{\sigma}{r}\right)^6
\right) +\varepsilon
\label{eq:wca}
\end{equation}
with the surface-to-surface distance $r$, cut-off radius $2^{1/6}\sigma$ and magnitude $\varepsilon$.

Based on the sum of the forces of \cref{eq:bbforce,eq:bbtorque,eq:consforce,eq:wca},
the swimmer's trajectory can be integrated using a standard symplectic Euler scheme.
We employ the waLBerla simulation framework\cite{godenschwager13a},
which implements the TRT LB algorithm of \cref{eq:lb}
and also includes a rigid-body integrator\cite{goetz10a}
and the moving boundaries of \cref{eq:vb,eq:bbforce,eq:bbtorque,eq:consforce}.
The same method has been implemented in previous works by other authors\cite{alarcon13a,alarcon17a,shen19a,shen18a,lintuvuori16a}.

\section{Far-field Results}
\label{sec:farfield}

In this section, we describe the simple validation tests for our numerical implementation.
We start with the far field, where we simulate the bulk flow field and interactions between two squirmers.
Next, we consider a squirmer confined in a narrow cylindrical tube,
and we conclude with the scattering of a squirmer off a spherical obstacle.

\subsection{Squirmers in Bulk}
\label{sec:bulk}

\begin{figure}[htb]
\centering
\includegraphics[width=\linewidth]{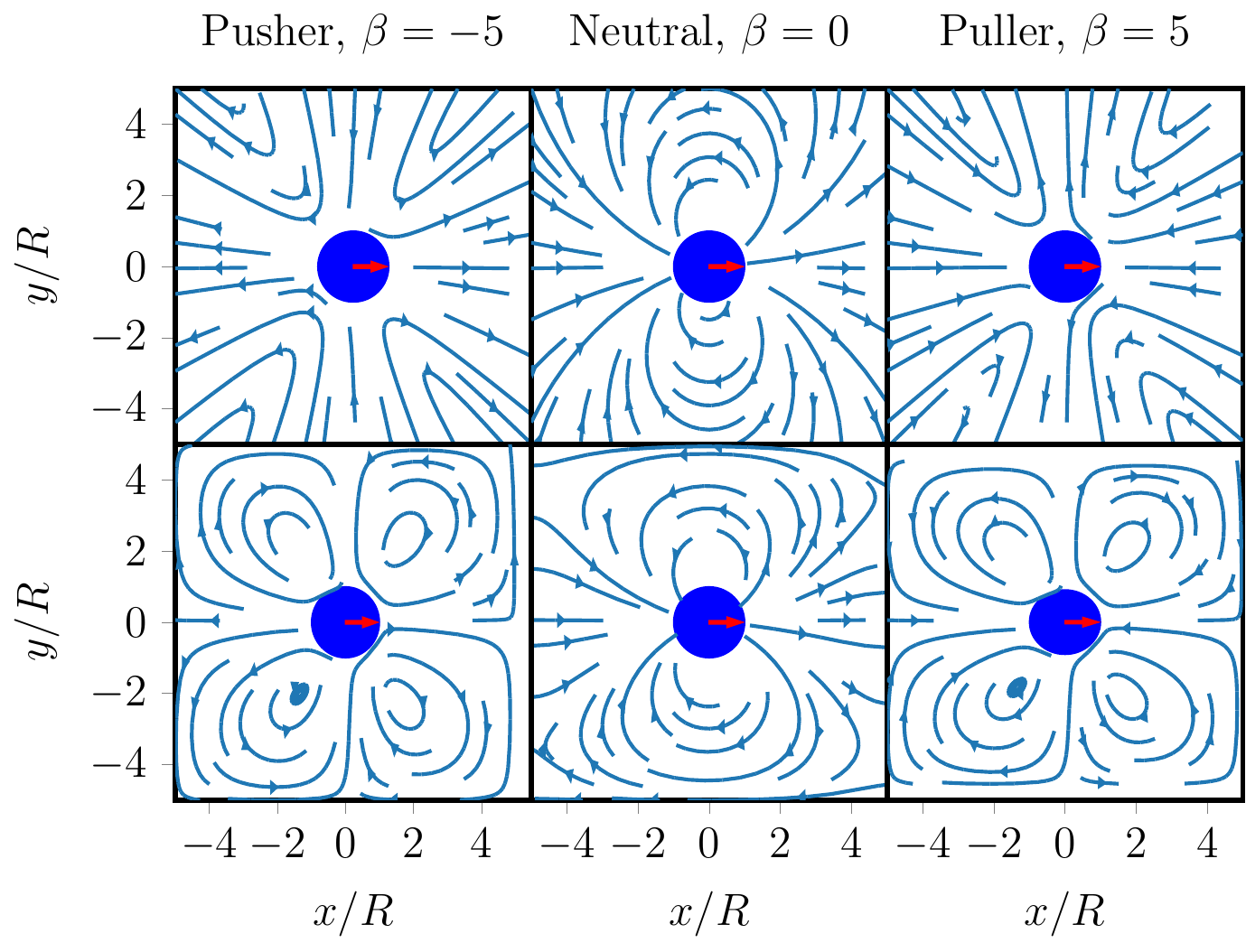}
\caption{Top: the analytical flow fields of squirmers with $\beta \in \{-5,0,5\}$ in an unbounded domain ($L=\infty$).
The red arrows indicate that the squirmer is oriented such that it moves to the right.
Bottom: the flow fields of the same squirmers at a resolution of $R=8$ as obtained via LB in a cubic box of length $L=10R$ with PBCs.
As one can see, the flow field is heavily influenced by the PBCs.
The analytical solution can also be determined for PBCs and looks indistinguishable from the LB flow fields.
}
\label{fig:bulk}
\end{figure}

In \cref{fig:bulk}, we show the flow fields of the three types of squirmer in bulk fluid.
As LB is typically used with periodic boundary conditions (PBCs),
a direct comparison to \cref{eq:squirmerflow} would require either an extremely large simulation domain in LB
or incorporating the effect of the periodic images into the analytical solution\cite{degraaf17b,adhyapak18a}.
The latter would require an Ewald summation approach\cite{brady88b,ishikawa08b},
but can be approximated by summing over a spherical shell of periodic images.
The largest differences between this approximate periodic analytical solution and the LB solution are found at $\pm 45^\circ$ from $\unitvec{e}$,
where the flow magnitude is small.
Ignoring these regions, the mean error is around $8\%$ at a resolution of $R=8$.
Both of these deviation can be attributed to discretization errors.
As we will discuss below, $R$ (in units of the lattic spacing) needs to have a certain minimal value to avoid more severe discretization artifacts.

\begin{figure}[htb]
\centering
\input{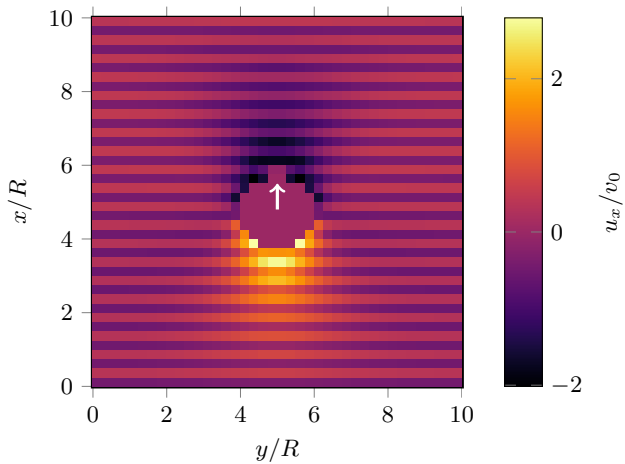}
\caption{Visualization of the deteriorated flow field after simulating a squirmer of radius $R=4$ in a box of length $L=10$ for $ T = 1.5\cdot10^5 \Delta t $. $ u_x $, the $ x $-component of the fluid velocity, is shown in the $ xy $-plane and normalized by the squirmer speed $v_0$. The arrow shown in grey indicates the squirmers orientation.}
\label{fig:stripes_color}
\end{figure}

\begin{figure}[htb]
\centering
\input{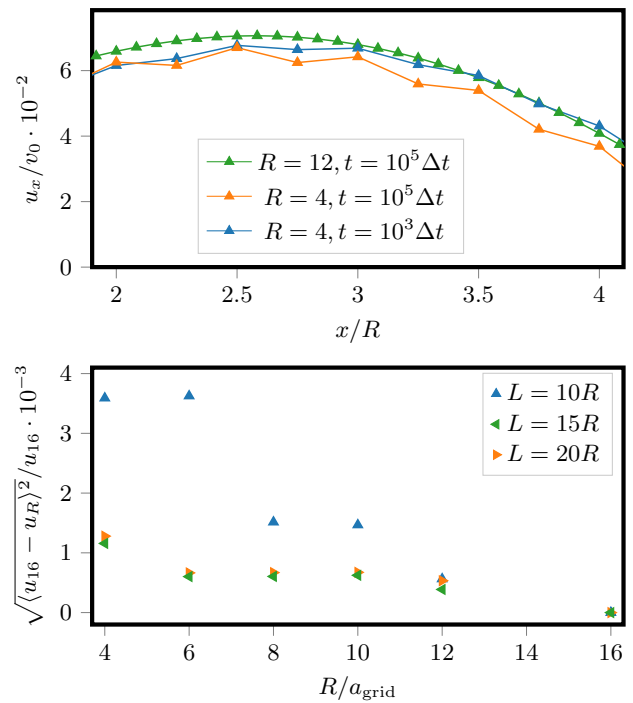}
\caption{Top: flow velocity $u_x(x\unitvec{e}_x)$, normalized by the squirmer speed $v_0$, along the $x$-axis for $L=10R$. For high resolutions ($R=12$), the curve is smooth, while for low resolutions ($R=4$) an alternating pattern of faster and slower cells is visible that grows more distinct over time, matching the stripes in \cref{fig:stripes_color}.
Bottom: Standard deviation of the velocity error obtained by comparing the flow field $u_R$ at a radius $R$ to that of a high-resolution simulation ($R=16$) for the entire simulation domain.
Larger values correspond to more inhomogeneous errors, i.e., the development of the alternating pattern described in the main text.
Time $t=1000$ and different box sizes $L\in \{10R,15R,20R\}$ are shown. 
The difference between $L=15R$ and $L=20R$ can be attributed to our error fitting procedure.}
\label{fig:stripes}
\end{figure}

\Cref{fig:bulk} was obtained at a resolution of 8 cells per squirmer radius.
Since we use lattice units, this corresponds to $R=8$.
In moving-boundary simulations of passive spheres,
one typically aims for a resolution of $R\approx 4$ which provides sufficient accuracy while minimizing computational effort\cite{nguyen02b}.
In literature, resolutions around $R=8$ are often used for squirmers\cite{shen19a, shen18a, lintuvuori16a}, but usually not explicitly justified.
Some authors\cite{alarcon13a, alarcon17a,pagonabarraga13a} do use smaller resolutions around $R=3$,
which for squirmers appears to only give usable results in the authors' specific case without preferred direction.
We find that resolutions below a value of $R\approx 6$ lead to strong oscillations in the flow field,
causing an alternating velocity pattern along the direction in which the squirmer moves, see \cref{fig:stripes_color}.
As seen in \cref{fig:stripes}, the magnitude of the oscillation increases over time,
suggesting a self-reinforcing numerical artifact.
While at short times, the true flow can still be obtained by averaging over the oscillation,
after several million time steps, they become so strong that the true flow is almost completely obscured.
Eventually the simulation becomes unstable because LB does not accurately handle strong velocity gradients like those in \cref{fig:stripes_color} well.
This phenomenon is most often seen in systems with a preferred direction.
\Citeauthor{alarcon17a}\cite{alarcon13a,alarcon17a}, for example, do not see this effect because they have dense suspensions of squirmers that continuously change their orientations.
For comparison, in MPCD, squirmer radii of three collision cells, each of which containing an average of 80 MPCD particles, are reported to have been used\cite{theers18a}.
Since the computational effort for an LB cell and for an MPCD particle are on the same order of magnitude,
the resolution requirement can be considered to be similar for LB and MPCD.

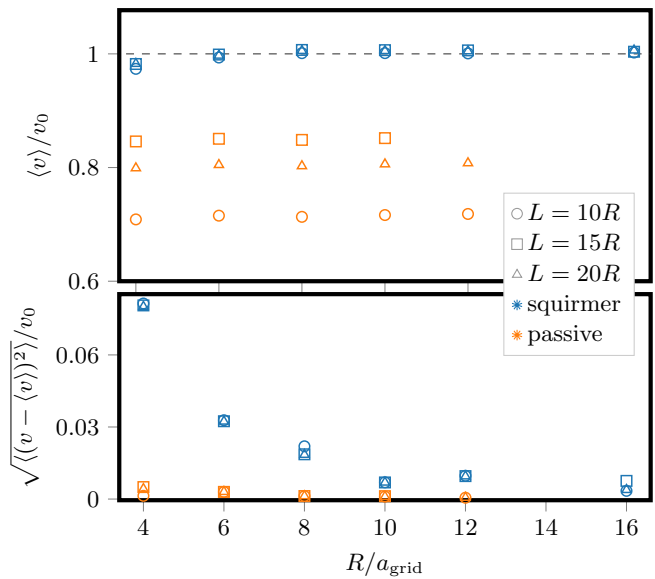
\begin{figure}[htb]
\centering
\begin{tikzpicture}

\definecolor{color0}{rgb}{0.12156862745098,0.466666666666667,0.705882352941177}
\definecolor{color1}{rgb}{1,0.498039215686275,0.0549019607843137}
\definecolor{color2}{rgb}{0.172549019607843,0.627450980392157,0.172549019607843}

\begin{axis}[axis line style = ultra thick,
height=.6\linewidth,
width=\linewidth,
tick align=outside,
tick pos=left,
xticklabels={,,},
x grid style={white!69.01960784313725!black},
xmin=3.6, xmax=16.4,
y grid style={white!69.01960784313725!black},
ylabel={$ \langle v\rangle/v_0 $},
ymin=.6, ymax=1.07687146623849
]

\addlegendimage{only marks, mark=*, mark size=2, mark options={solid,rotate=0 ,fill opacity=0}, black!40!}
\addlegendimage{only marks, mark=square  , mark size=2, mark options={solid           ,fill opacity=0}, black!40!}
\addlegendimage{only marks, mark=triangle, mark size=2, mark options={solid,rotate=0,fill opacity=0}, black!40!}
\addlegendimage{only marks, mark=10-pointed star, mark size=2, mark options={solid,rotate=0,fill opacity=100}, color0}
\addlegendimage{only marks, mark=10-pointed star, mark size=2, mark options={solid,rotate=0,fill opacity=100}, color1}


\addplot [line width=0.5599999999999999pt, color0, mark=*, mark size=2, mark options={solid,rotate=0,fill opacity=0}, only marks, forget plot]
table [row sep=\\]{%
	4	0.97365650212793 \\
	6	0.99355365570849 \\
	8	1.00125115603515 \\
	10	1.00133163799956 \\
	12	1.000816395557 \\
	16	1.00237056023524 \\
};
\addplot [line width=0.5599999999999999pt, color0, mark=triangle, mark size=2, mark options={solid,rotate=0,fill opacity=0}, only marks, forget plot]
table [row sep=\\]{%
	4	0.982698707847231 \\
	6	0.997144670909674 \\
	8	1.00549716543689 \\
	10	1.00531560959017 \\
	12	1.00484611736834 \\
	16	1.0062420693568 \\
};
\addplot [line width=0.5599999999999999pt, color0, mark=square, mark size=2, mark options={solid,fill opacity=0}, only marks, forget plot]
table [row sep=\\]{%
	4	0.982189445779954 \\
	6	0.998807610613665 \\
	8	1.00691845858635 \\
	10	1.00656623194654 \\
	12	1.00611180204596 \\
	16	1.00376789661413 \\
};
\draw[black!80!, dashed] (axis cs:3,1) -- (axis cs: 17,1);
%
%
\addplot [line width=0.5599999999999999pt, color1, mark=*, mark size=2, mark options={solid,rotate=0,fill opacity=0}, only marks, forget plot]
table [row sep=\\]{%
	10	0.7164256685 \\
	12	0.718336047888889 \\
	4	0.708744658 \\
	6	0.715330772555556 \\
	8	0.713252481055556 \\
};
\addplot [line width=0.5599999999999999pt, color1, mark=triangle, mark size=2, mark options={solid,rotate=0,fill opacity=0}, only marks, forget plot]
table [row sep=\\]{%
	10	0.805933151944444 \\
	12	0.807859508922881 \\
	4	0.799010956 \\
	6	0.804719341722222 \\
	8	0.8026489385 \\
};
\addplot [line width=0.5599999999999999pt, color1, mark=square, mark size=2, mark options={solid,fill opacity=0}, only marks, forget plot]
table [row sep=\\]{%
	10	0.851767244600117 \\
	4	0.845933784 \\
	6	0.850590929277778 \\
	8	0.848652713777778 \\
};
\end{axis}

\begin{scope}[yshift=-.337\linewidth]
\begin{axis}[axis line style = ultra thick,
height=.5\linewidth,
width=\linewidth,
legend cell align={left},
legend entries={{$L = 10R$},{$L = 15R$},{$L = 20R$},{squirmer},{passive}},
legend style={at={(0.97,.7)}, anchor=south east, draw=white!80.0!black},
tick align=outside,
tick pos=left,
x grid style={white!69.01960784313725!black},
xlabel={$R/a_\text{grid}$},
xmin=3.4, xmax=16.6,
y grid style={white!69.01960784313725!black},
ylabel={$ \sqrt{\langle (v -\langle v\rangle)^2 \rangle}/v_0 $},
ymin=-0.000478863261294591, ymax=0.0852955086382198,
ytick={0,0.03,0.06},
yticklabel style={/pgf/number format/fixed,/pgf/number format/precision=2},
scaled y ticks=false
]

\addlegendimage{only marks, mark=*, mark size=2, mark options={solid,rotate=0 ,fill opacity=0}, black!40!}
\addlegendimage{only marks, mark=square  , mark size=2, mark options={solid           ,fill opacity=0}, black!40!}
\addlegendimage{only marks, mark=triangle, mark size=2, mark options={solid,rotate=0,fill opacity=0}, black!40!}
\addlegendimage{only marks, mark=10-pointed star, mark size=2, mark options={solid,rotate=0,fill opacity=100}, color0}
\addlegendimage{only marks, mark=10-pointed star, mark size=2, mark options={solid,rotate=0,fill opacity=100}, color1}

\addplot [line width=0.5599999999999999pt, color0, mark=*, mark size=2, mark options={solid,rotate=0,fill opacity=0}, only marks]
table [row sep=\\]{%
4	0.0813966735518782 \\
6	0.0328120645920352 \\
8	0.0219588133434952 \\
10	0.0069663803717201 \\
12	0.00968837822290951 \\
16	0.00341997182504697 \\
};
\addplot [line width=0.5599999999999999pt, color0, mark=triangle, mark size=2, mark options={solid,rotate=0,fill opacity=0}, only marks]
table [row sep=\\]{%
4	0.0800935218269843 \\
6	0.0323283558027371 \\
8	0.0184826626837666 \\
10	0.00693202623154326 \\
12	0.0095057490428249 \\
16	0.00384440975051608 \\
};
\addplot [line width=0.5599999999999999pt, color0, mark=square, mark size=2, mark options={solid,fill opacity=0}, only marks]
table [row sep=\\]{%
4	0.0806058667639028 \\
6	0.0323946730632981 \\
8	0.0186366173841222 \\
10	0.00691519413953776 \\
12	0.00955469017153618 \\
16	0.00756569476910685 \\
};


\addplot [line width=0.5599999999999999pt, color1, mark=*, mark size=2, mark options={solid,rotate=0,fill opacity=0}, only marks, forget plot]
table [row sep=\\]{%
10	0.00104457055271903 \\
12	0.000552422174745998 \\
4	0.00135319964862395 \\
6	0.00297176037729269 \\
8	0.00123907695814129 \\
};
\addplot [line width=0.5599999999999999pt, color1, mark=triangle, mark size=2, mark options={solid,rotate=0,fill opacity=0}, only marks, forget plot]
table [row sep=\\]{%
10	0.00103311860057659 \\
12	0.000610991495522181 \\
4	0.00424219676465673 \\
6	0.00302870087348673 \\
8	0.00125812495225853 \\
};
\addplot [line width=0.5599999999999999pt, color1, mark=square, mark size=2, mark options={solid,fill opacity=0}, only marks, forget plot]
table [row sep=\\]{%
10	0.00121058232260436 \\
4	0.00498707461196882 \\
6	0.00298268190207116 \\
8	0.00125119146153184 \\
};

\end{axis}
\end{scope}

\end{tikzpicture}
\caption{Top: A squirmer's (orange) and a passive sphere's (blue) mean velocity $\langle v \rangle$, normalized by the expected bulk speed $v_0$, as a function of the resolution of the squirmer/passive sphere.
Bottom: Normalized variance of the respective bodies velocities indicating the spread around the mean velocity. The resolution is given by the radius $R$ in lattice units.
We show results for three box sizes as given by the edge length $L$, also in lattice units.}
\label{fig:resolution}
\end{figure}

In \cref{fig:resolution}, we show how the squirmer's speed $v$ depends on the resolution as given by the squirmer's radius $R$ and on the box length $L$.
In analytical theory and in LB in the infinite-resolution limit, this speed equals the squirmer parameter $v_0$ from \cref{eq:squirmer-speed}.
In \cref{fig:resolution} we observe that we approach $v_0$ from below as resolution increases.
At $R=6$, we are already within $0.5\%$ of the correct value ($0.2\%$ at $R=12$) for the largest box size.
For the smaller box sizes,
the interaction of the squirmer with its periodic images decreases the velocity slightly.
For comparison, the same data is also plotted for a passive sphere being dragged through a resting fluid at otherwise identical parameters.
The periodicity effect is much weaker for the squirmer than for the passive sphere since the latter's flow is monopolar to leading order and thus decays more slowly than the squirmer's.
Despite the seemingly good agreement of the observed squirmer velocity with the prescribed squirmer velocity even for small resolutions,
the stripe pattern discussed in the previous paragraph massively modifies the flow field, to the extent that simulations at small resolutions simply give no meaningful results.

While the squirmer moves across the lattice,
some variation in its speed is expected
due to the sphere being composed of discrete cubes.
As expected, \cref{fig:resolution} shows that the variation decreases with resolution,
however the variation is much larger than for the equally-resolved passive sphere dragged through the fluid.
The latter can be attributed to the cause of the motion---the squirmer is dragged along by the flow its own surface causes---and to the fact that this surface is significantly affected by the slight changes in the number of cells occupied by a sphere as it moves.
For a graphical explanation of this problem, see Ref.~\citenum{kuron16a},
where we introduced a solution to a related problem for an electrophoretically-driven particle.
For the squirmer, an approach based on the method of \citet{noble98a} might prove useful.

\subsection{Interaction between Two Squirmers}
\label{sec:passing}

\begin{figure}[htb]
\centering
\input{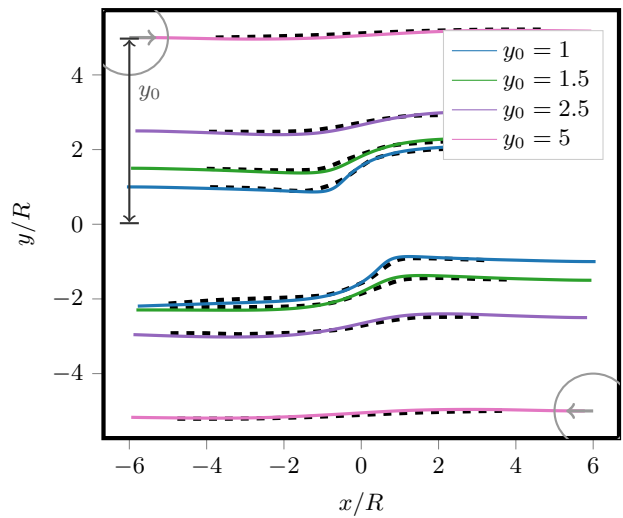}
\caption{Trajectories of two squirmers with $\beta=5$ passing each other. The initial configuraiton is specified by the initial lateral separation $d = 2y_0$ and the separation in the direction of their original orientation.
The results of Ref.~\citenum{ishikawa06a} are shown as dashed lines.
Our LB simulations are performed for squirmers with $ R = 9 $ in a periodic box of length $L=250$.}
\label{fig:ishikawa}
\end{figure}

Now that we have confirmed that our LB implementation is capable of producing the correct flow around a squirmer,
we check that a squirmer correctly reacts to the flow produced by another squirmer.
Here, we approximated the situation first considered by \citet{ishikawa06a},
who positioned two squirmers facing opposite each other, separated by a distance of $12R$ and spaced apart laterally by varying distances $d$.
They solved for the trajectories using BEM,
which assumes an infinite fluid domain size and discretizes the squirmers' surfaces.
In our LB calculations, we used a squirmer radius of $R=9$
and a cubic, periodic simulation domain with edge length $L=250$
to approximate the bulk calculation of Ref.~\citenum{ishikawa06a}.
\Cref{fig:ishikawa} shows the resulting trajectories
and those of Ref.~\citenum{ishikawa06a}.
There is good agreement and our trajectories are considerably smoother than those given by \citeauthor{ishikawa06a}.
Part of the deviations can be attributed to the use of PBCs in our simulation\cite{degraaf17a};
the increased smoothness is mostly related to the advancement in computational performance since \citeyear{ishikawa06a}
and not an intrinsic issue with BEM.
It is worth noting that we have used a far coarser resolution for our squirmers than used by \citeauthor{ishikawa06a}
because we also need to discretize the entire fluid volume,
while BEM only discretizes the surface.
The good match between the much finer BEM resolution and our LB results is promising for simulations at much higher squirmer volume fractions,
where we can reasonably expect to be able to maintain our current resolution
and thus keep roughly the same simulation speed.

\section{Near-field Results}
\label{sec:nearfield}

Now that we have determined the resolution required for an accurate simulation
and checked that two squirmers interact with each other correctly,
we can investigate systems where the near field plays a dominant role.

\subsection{Squirmer in a Round Tube}
\label{sec:channel}

LB does not make use of the method of reflections to capture the effect of solid/no-slip boundaries
and therefore can accurately reproduce the near-field flow when squirmers approach obstacles closer than their diameter.
Note we do not include lubrication corrections\cite{ishikawa08b,nguyen02b}, which would be necessary to accurately capture the flow between two objects that are spaced less than one lattice constant apart.
Keeping this in mind, we can now perform simulations where the near-field flow plays a role.
\citet{zhu13a} study a neutral ($\beta=0$) squirmer oscillating in a tube with circular diameter $ D = 20R/3 $ and length $ L = 3\pi R $ with PBCs only along its length, starting from different distances $y_0$ to the boundary and an initial orientation parallel to the symmetry axis.

\begin{figure}[htb]
\centering
\input{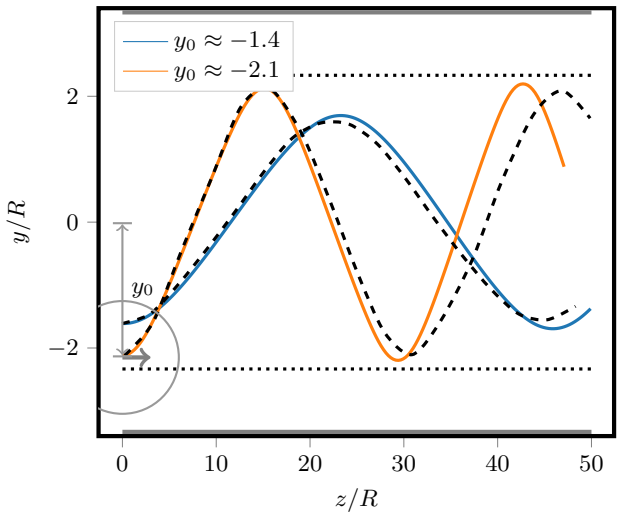}
\caption{Trajectories of squirmers ($R=9$) with different initial displacements $ y_0 $ inside a tube of length $ L = 3\pi R $ and circular diameter $ D = 20R/3 $, compared to the results of Ref.~\citenum{zhu13a}, which simulates the equivalent system using BEM.
These are shown using black dashes. The dotted lines indicate the point of closest approach before the squirmer touches the boundary, which itself is marked using the grey lines at the top and bottom of the plot. Also shown: the location of the squirmer and its initial orientation in the case of $ y_0 \approx -2.1 $.}
\label{fig:tube}
\end{figure}

The trajectories we obtain are compared to the results from that publication in \cref{fig:tube}.
Both the oscillation amplitude and the period match to within $3\%$ of the literature value.
The reorientation of the squirmer when it is near the wall is a near-field effect,
so the agreement confirms that our LB method sufficiently captures it.
Since \citet{zhu13a} use BEM with local mesh refinement\cite{ishikawa06a,ishikawa06b} when squirmer and wall are near contact,
they capture near-field effects more accurately than LB does at the resolution we used.
This explains the slight deviations in the trajectories of \cref{fig:tube},
but the good agreement confirms that the system is rather robust to these differences.
To obtain equally good results in LB at manageable computational effort,
one would need to resort to an adaptive grid resolution\cite{schornbaum16a}.

\subsection{Scattering off and Orbiting around a Stationary Spherical Obstacle}
\label{sec:orbiting}

The last system we consider is that of a single squirmer scattering off or orbiting around a stationary spherical obstacle with radius $A$.
\citet{spagnolie15a} have examined how force dipole swimmers interact with such an obstacle
and found that, depending on the value of $\beta$ and the ratio $A/R$, the swimmer may either orbit around or scatter off the obstacle.
Here, we compare to their results using our LB squirmer.
However, there are some notable differences between our method
and the calculation by \citeauthor{spagnolie15a}
which we will briefly address in the following.

The force dipole is the slowest-decaying mode contained in the squirmer (\cref{eq:squirmerflow}),
so the far field agrees between our and their point-dipole approximation.
However, close to the obstacle a far-field description is not sufficient
to account for the finite size of a microswimmer.
In our LB squirmer simulations, we do account for this effect,
so deviations are expected with respect to the result of Ref.~\citenum{spagnolie15a}.
Even if $\left|\beta\right|$ is chosen sufficiently large for the force dipole to dominate,
qualitative agreement is only expected to a certain extent:
since squirmers also contain a quadrupolar contribution, see \cref{eq:squirmerflow},
their behavior near surfaces is altered compared to a pure dipole\cite{degraaf16b,mathijssen15a,shen17a}.
In addition to this hydrodynamic effect,
there is a difference in the contact potential:
We use a short-ranged WCA repulsion (\cref{eq:wca}),
while Ref.~\citenum{spagnolie15a} uses a hard-core repulsion;
the effect of this should be less pronounced than that of the addition of higher-order hydrodynamic moments.
Finally, we would like to note that
\citet{chamolly17a} study a similar system, using actual squirmers instead of dipoles and a short-range repulsion that is softer than our WCA.
However, they use SD\cite{ishikawa08b,brady88a}, a far-field-only hydrodynamics solver (for $d\gg R$),
combined with lubrication corrections\cite{ishikawa08b,durlofsky87a} (for $d\ll R$), meaning that near-field interactions ($\mathcal{O}(d)\approx\mathcal{O}(R)$) are not dealt with.
Because \citeauthor{chamolly17a} only study squirmers in high-volume fraction periodic crystals, we have excluded their results from our comparison.

\begin{figure}[htb]
\centering
\begin{tikzpicture}

\definecolor{color0}{rgb}{0.12156862745098,0.466666666666667,0.705882352941177}
\definecolor{color1}{rgb}{1,0.498039215686275,0.0549019607843137}
\definecolor{color2}{rgb}{0.172549019607843,0.627450980392157,0.172549019607843}

\begin{axis}[axis line style=ultra thick,
legend cell align={left},
legend entries={{Dipole reference},{Squirmer simulation}},
legend style={draw=white!80.0!black, fill=white},
tick align=outside,
tick pos=left,
x grid style={white!69.01960784313725!black},
xlabel={$-\beta$},
xmin=0.2, xmax=45.5,
xmode=log,
xtick={0.01,0.1,1,10,100,1000},
xticklabels={$10^{-2}$,$10^{-1}$,$10^{0}$,${10^{1}}$,${10^{2}}$,${10^{3}}$},
y grid style={white!69.01960784313725!black},
ylabel={critical $A/R$},
ymin=-2.4524, ymax=189.0004,
ymode=log,
ytick={0.1,1,10,100,1000,10000,100000},
yticklabels={${10^{-1}}$,${10^{0}}$,${10^{1}}$,$10^{2}$,${10^{3}}$,${10^{4}}$,${10^{5}}$}
]
\addlegendimage{mark=*, mark size=2, mark options={solid}, color0}
\addlegendimage{mark=*, mark size=2, mark options={solid}, only marks, color1}
\addplot [line width=0.5599999999999999pt, color0, mark=*, mark size=2, mark options={solid}, only marks]
table [row sep=\\]{%
	0.266666666666667	181.201 \\
	0.400924	82.116 \\
	0.532717333333333	48.824 \\
	0.532889333333333	48.803 \\
	0.666974666666667	32.768 \\
	0.801361333333333	25.536 \\
	0.933270666666667	18.921 \\
	1.06479066666667	16.418 \\
	1.19747466666667	13.916 \\
	1.33233333333333	11.88 \\
};
\addplot [line width=0.5599999999999999pt, color1, mark=|, mark size=5, mark options={solid}, only marks, forget plot]
table [row sep=\\]{%
32	6.25 \\
14	9 \\
13	10.5 \\
10	12 \\
};
\addplot [line width=0.5599999999999999pt, color1, mark=|, mark size=5, mark options={solid}, only marks, forget plot]
table [row sep=\\]{%
34	6.25 \\
16	9 \\
15	10.5 \\
14	12 \\
};
\addplot [line width=0.5599999999999999pt, color1, mark=*, mark size=2, mark options={solid}, only marks]
table [row sep=\\]{%
33	6.25 \\
15	9 \\
14	10.5 \\
12	12 \\
};

\path [draw=color1, line width=0.5599999999999999pt] (axis cs:32,6.25)
--(axis cs:34,6.25);

\path [draw=color1, line width=0.5599999999999999pt] (axis cs:14,9)
--(axis cs:16,9);

\path [draw=color1, line width=0.5599999999999999pt] (axis cs:13,10.5)
--(axis cs:15,10.5);

\path [draw=color1, line width=0.5599999999999999pt] (axis cs:10,12)
--(axis cs:14,12);

\addplot[domain=.19:1.4, color0] {1024/(81*x^2)};

\end{axis}

\begin{scope}[scale=.5, xshift=1.2\linewidth, yshift=6cm]
\draw[draw=black!40!] (-4, 0) -- (0,0);
\draw[black] (-4.2,2) rectangle (2.7,-2);
\draw[fill=black!40!white!] (0,0) circle (1);
\draw[|<->|] (0,0) -- node[above, sloped] (r) { $A$ } (20:.9) --+ (20:.1);

\node (s) at (-3.7, .5) {};
    \draw[->, thick] (s) -- node[above, yshift=-.5] {$\unitvec{e}$}  + (1,0);
\draw[fill=red!50!white] (s) circle (.2);
\draw[|<->|] (s)+(-.2, .4) -- node[above] (R) {$ 2R $} + (.2, .4);
\draw[|<->|] (-2, .5) -- node[right] {$ y_0 $} (-2,0);

\end{scope}

\end{tikzpicture}
\caption{Critical obstacle size $A/R$ below which the squirmer scatters and above which it orbits, as a function of $\beta$, including error bars. The data shown in blue is extracted from Ref.~\citenum{spagnolie15a} for force dipole pushers; the blue line is an analytical solution given there for small $\left|\beta\right|$. The inset schematically shows the system under investigation: a squirmer of radius $ R $ swims with a small offset $ y_0 $ above the $ x $-axis towards the spherical obstacle of radius $ A $.}
\label{fig:orbit}
\end{figure}
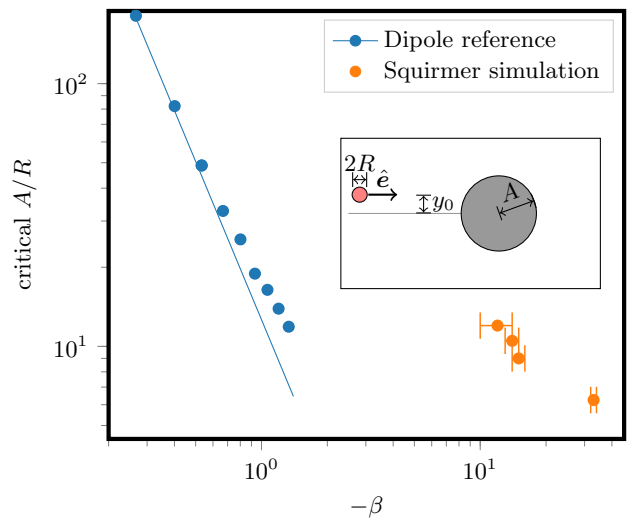

In \cref{fig:orbit}, we present the critical obstacle sizes $A/R$ for different $\beta$;
below this curve the squirmers scatter and above they orbit.
Reference data from Ref.~\citenum{spagnolie15a} is also shown,
but needs to be considered with care since the model differs significantly in the ways discussed above.
Furthermore, our LB model is only capable of relatively small $A/R$ ratios
as a constant resolution is used throught the simulation domain,
which makes the computational effort scale like $\mathcal{O}(A^3)$.
To extend our results into the realm of parameters studied by \citeauthor{spagnolie15a},
an adaptive grid resolution\cite{schornbaum16a} would be helpful.
To obtain our figure, simulations at various values $A$ and $\beta$ are started with the squirmer of radius $ R = 8 $ positioned such that it moves radially toward the obstacle.
In practice, one needs to break the symmetry by angling the trajectory slightly instead of using a perfectly radial one.
While such an offset angle influences the angle by which a squirmer is scattered by the obstacle\cite{spagnolie15a}, it does not affect the critical $A/R$, i.e., whether a squirmer enters into an orbit or not.
A trajectory is considered to be an orbit if the squirmer revolves more than halfway around the obstacle,
though we have performed simulations of a select number of situations to confirm that the squirmer indeed completes a full revolution and continues to orbit.
To determine the exact position of the critical value,
we performed bisection in $A/R$ or $\beta$ once at least one orbiting and one scattering parameter set had been found.

\section{Conclusion and Outlook}
\label{sec:conclusion}

We have described in detail our implementation of a hydrodynamic squirmer model\cite{blake71b,lighthill52a} in an LB fluid dynamics solver,
where we build upon a large body of literature on this topic\cite{mcnamara88a,aidun98a,ladd94a}.
We have confirmed in four scenarios that our LB squirmer implementation can accurately reproduce signature features of the squirmer model, including:
(i) The analytic flow field around the squirmer,
accounting for periodicity effects.
(ii) The interaction between two squirmers, as originally obtained by \citet{ishikawa06a} using the boundary element method (BEM).
(iii) The oscillation of a squirmer in a cylindrical tube as studied originally using BEM by \citet{zhu13a}.
(iv) The scattering and orbiting of a squirmer around a spherical obstacle.
This problem was analyzed theoretically by \citet{spagnolie15a} for a point-like
force-dipole swimmer.
Our squirmer model can only probe smaller size ratios,
but yields qualitatively similar trends of scattering vs. orbiting in a size regime beyond theirs.

Through our study we have also demonstrated that the LB squirmer implementation is sensitive to discretization artifacts,
more so than has been reported for passive particles.
Throughout the literature various values of the resolution of the squirmer are used.
Here, we show that a refinement of at least 8 lattice cells for the radius of the squirmer is necessary to avoid numerical artifacts in the flow field.
These artifacts are particularly pronounced in situations where there is persistent motion and may lead to severe numerical instabilities.



We have made our LB squirmer implementation available within the open-source software waLBerla\cite{godenschwager13a},
which will make it possible for anyone to simulate large-scale systems
containing many squirmers and complex boundary conditions.
It should be noted that we have not incorporated lubrication corrections\cite{nguyen02b,ishikawa08b} here,
which will be a topic for further method development.


\begin{acknowledgments}
  We thank the Deutsche Forschungsgemeinschaft (DFG) for funding
  through the SPP 1726 ``Microswimmers: from single particle motion to
  collective behavior'' (HO1108/24-1 and HO1108/24-2).
  We are grateful to Alexander Chamolly for useful discussions
  and to Martin Bauer, Sebastian Eibl, Christian Godenschwager, Christoph Rettinger, and Florian Schornbaum
  for developing waLBerla and supporting us in using and extending it.
\end{acknowledgments}


\section*{References}
\raggedright
\bibliographystyle{jabbrv_apsrev4-1}
\bibliography{bibtex/icp}

\end{document}